\documentclass[12pt]{article}
\usepackage{epsfig}
\usepackage{amssymb}
\usepackage{amsmath}
\usepackage{bm}                  
\usepackage{cancel}
\usepackage{comment} 
\newcommand{\pres}[2]{\setbox0=\hbox{$\scriptstyle #1$} \dimen0=\dp0  
              \dimen1=\ht0 \divide\dimen1 by 3
              \advance\dimen0 by \dimen1
              \hbox{\lower\dimen0 \hbox{$\scriptstyle #2\!\!$}} #1}
\newcommand{\nc}{\newcommand}
\nc{\figcap}[1]{\begin{quote}\refstepcounter{figure}
        {\bf Figure \thefigure}: {\small #1}\end{quote}}
\nc{\fig}[1]{\mbox{Fig.~\ref{#1}}}
\nc{\noi}{\noindent}
\nc{\bea}{\begin{eqnarray}}
\nc{\eea}{\end{eqnarray}}
\nc{\bean}{\begin{eqnarray*}}
\nc{\eean}{\end{eqnarray*}}
\nc{\ba}{\begin{array}}
\nc{\ea}{\end{array}}
\nc{\be}{\begin{equation}}
\nc{\ee}{\end{equation}}
\nc{\nn}{\nonumber}
\nc{\bra}[1]{\langle #1|}
\nc{\ket}[1]{|#1\rangle}
\nc{\av}[1] {\langle #1\rangle}
\nc{\vac}[1] {\langle 0| #1|0\rangle}
\nc{\amp}[2]{\langle #1|#2\rangle}
\nc{\da}{\dagger}
\nc{\pa}{\partial}
\nc{\ga}{\gamma}
\nc{\ep}{\epsilon}
\nc{\tf}{t_f}
\nc{\half}{\ensuremath{\frac{1}{2}}}
\nc{\hHH}{\hat H}
\nc{\ha}{\hat a}
\nc{\hO}{\hat O}
\nc{\hAA}{\hat A}
\nc{\hB}{\hat B}
\nc{\hG}{\hat G}
\nc{\hN}{\hat N}
\nc{\hU}{\hat U}
\nc{\hx}{\hat{x}}
\nc{\hp}{\hat{p}}
\nc{\hpsi}{\hat \psi}
\nc{\hphi}{\hat \phi}
\nc{\hpi}{\hat \pi}
\nc{\hpd}{\hat \psi ^\dagger}
\nc{\hE}{\hat E}
\nc{\hb}{\hat b}
\nc{\hc}{\hat c}
\nc{\hjo}{\hat j _0}
\nc{\hrho}{\hat \rho}
\nc{\leave}{\! \! \! \! \! / \, \,}
\nc{\intl}[1]{\int d\! #1 \,} 
\nc{\intll}[3]{\int _#1^#2 d\! #3 \,} 
\nc{\lm}{\lim _{y \rightarrow x}}
\nc{\scd}{\partial ^2 _{A_T}}
\nc{\fd}[1]{\frac{\delta }{\delta #1}} 
\nc{\pad}[1]{\frac{\partial}{\partial #1}} 
\nc{\refpa}[1]{(\ref{#1})} 
\nc{\calH}{\ensuremath{\mathcal{H}}}
\nc{\calD}{\ensuremath{\mathcal{D}}}
\nc{\calL}{\ensuremath{\mathcal{L}}}
\nc{\calO}{\ensuremath{\mathcal{O}}}
\nc{\hcalO}{\ensuremath{\hat \mathcal{O}}}
\nc{\calK}{\ensuremath{\mathcal{K}}}
\nc{\Tr}{\ensuremath{\mathrm{Tr}}}
\nc{\tr}{\ensuremath{\mathrm{tr}}}
\nc{\ra}{\rightarrow}
\nc{\lr}{\leftrightarrow}
\nc{\phistar}{\phi^*}
\nc{\etat}{\eta_T}
\nc{\het}{\hat E_T}
\nc{\hpt}{\hat \psi_T}
\nc{\hpdt}{\hat \psi ^\dagger_T}
\nc{\bart}{\bar{t}}
\nc{\barp}{\bar{p}}
\nc{\barT}{\bar{T}}
\nc{\hbarrho}{\hat{\bar{\rho}}}
\nc{\bga}{\ensuremath{\mbox{\boldmath{$\gamma$}}}}
\nc{\bsi}{\ensuremath{\mathbf{\sigma}}}
\nc{\bx}{\ensuremath{\mathbf{x}}}
\nc{\by}{\ensuremath{\mathbf{y}}}
\nc{\bz}{\ensuremath{\mathbf{z}}}
\nc{\bp}{\ensuremath{\mathbf{p}}}
\nc{\bn}{\ensuremath{\mathbf{n}}}
\nc{\bbp}{\ensuremath{\bar{\mathbf{p}}}}
\nc{\bP}{\ensuremath{\mathbf{P}}}
\nc{\hbA}{\hat{\ensuremath{\mathbf{A}}}}
\nc{\hbB}{\hat{\ensuremath{\mathbf{B}}}}
\nc{\bA}{\ensuremath{\mathbf{A}}}
\nc{\bJ}{\ensuremath{\mathbf{J}}}
\nc{\bB}{\ensuremath{\mathbf{B}}}
\nc{\bH}{\ensuremath{\mathbf{H}}}
\nc{\bM}{\ensuremath{\mathbf{M}}}
\nc{\bD}{\ensuremath{\mathbf{D}}}
\nc{\bE}{\ensuremath{\mathbf{E}}}
\nc{\hbE}{\hat{\ensuremath{\mathbf{E}}}}
\nc{\br}{\ensuremath{\mathbf{r}}}
\nc{\bj}{\ensuremath{\mathbf{j}}}
\nc{\bOm}{\ensuremath{\mathbf{\Om}}}
\nc{\om}{\omega}
\nc{\Om}{\Omega}
\nc{\sgn}{\mbox{sgn}}
\nc{\deltabar}{\mbox{$\delta\hspace*{-8pt}\vspace*{-8pt}-$}}
\nc{\gammat}{\tilde{\gamma}}
\nc{\mub}{\bar{\mu}}
\nc{\rhob}{\bar{\rho}}
\nc{\Bb}{\bar{B}}
\nc{\Jb}{\bar{J}}
\nc{\Mb}{\bar{M}}
\nc{\Tb}{\bar{T}}
\nc{\sbar}{\bar{s}}
\nc{\betab}{\bar{\beta}}
\nc{\hj}{\hat j}
\nc{\hQ}{\hat Q}
\nc{\hJ}{\hat J}
\nc{\hA}{\hat A}
\nc{\hH}{\hat H}
\nc{\de}{\delta}
\nc{\leri}{\leftrightarrow}
\nc{\llabel}[1]{\label{#1}\marginpar{#1}}
\nc{\bc}{\begin{center}}
\nc{\ec}{\end{center}}
\nc{\inv}[1]{\frac{1}{#1}}

\newlength{\overeqskip}
\newlength{\undereqskip}
\nc{\eq}[1]{\mbox{Eq.~(\ref{#1})}}
\nc{\eps}{\epsilon}
\nc{\goto}{\rightarrow}
\nc{\cF}{{\cal F}}
\nc{\cG}{{\cal G}}
\nc{\cH}{{\cal H}}
\newcounter{sectionc}
\newcounter{subsectionc}
\newcounter{subsubsectionc}
\renewcommand{\section}[1]
{\refstepcounter{sectionc}\vspace{0.0cm}
\setcounter{subsectionc}{0}\setcounter{subsubsectionc}{0}\noindent 
	{\bf\thesectionc. #1}}
\renewcommand{\subsection}[1] {\vspace{0.0cm}
\addtocounter{subsectionc}{1} 
	\setcounter{subsubsectionc}{0}\noindent 
	{\it\thesectionc.\thesubsectionc. #1}\par\vspace{0.4cm}}
\renewcommand{\subsubsection}[1] {\vspace{0.6cm}\addtocounter{subsubsectionc}{1}
	\noindent {\rm\thesectionc.\thesubsectionc.\thesubsubsectionc. 
	#1}\par\vspace{0.0cm}}

\renewenvironment{thebibliography}[1]
	{\begin{list}{[\arabic{enumi}]}
	{\usecounter{enumi}\setlength{\parsep}{0pt}
\setlength{\leftmargin 0.52cm}{\rightmargin 0pt}
	 \setlength{\itemsep}{6pt} \settowidth
	{\labelwidth}{#1.}\sloppy}}{\end{list}}
\newcommand{\seqnoll}{\setcounter{equation}{0}}
\begin{document}
%
%
%
%
%
\setlength{\jot}{10pt} 
%
%
\thispagestyle{empty} 
%
%
%
%
\vspace*{10mm}
\begin{center}  
\baselineskip 1.5cm 
{
\Large\bf
Comments on the Three-Slit Experiment and Quantum Mechanics
}
\\[5mm]  
\normalsize 
\end{center} 
\begin{center} 
\vspace*{5mm}
{\centering 
{\large Bo-Sture K. Skagerstam\footnote{Email address: bo-sture.skagerstam@ntnu.no}
\vspace*{10mm}
\\
Department of Physics \\  Norwegian University of Science and Technology, NTNU \\ N-7491 Trondheim,  Norway}}
%
%
%
%
\end{center} 
%
%
%
%
%
%
%
%
%
%
\vspace*{10mm}
\begin{abstract} 
%
%
\noindent It has been suggested by Sorkin that a three-slit Young experiment could reveal the validity a fundamental  ingredient in the foundations of one of the cornerstones in modern physics namely quantum mechanics. In terms of a certain parameter $\kappa_S$, it was argued that a non-zero value could imply a breakdown of the fundamental Born's rule as well as the superposition principle.  Here we argue that a  physical realization of such  arguments  could lead to an erroneous conclusion and  contradict the basic rules of quantum mechanics. In fact, we argue that a proper interpretation of the procedures involved in a  physical  determination of $\kappa_S$ does not necessarily lead to $\kappa_S=0$. In order to show this we consider a mono-chromatic source of photons prepared in an {\it arbitrary} quantum state and  a simple version of the well-established photon detection theory of Glauber which, by construction,  obeys all the rules of quantum mechanics. It is, however,  also argued that after a proper identification of the relevant quantum-mechanical probability amplitudes one can be reach  $\kappa_S=0$.
As long as one only consider a single photon detector, it is verified that, in this context, there is no fundamental difference between quantum-mechanical interference and interference as expressed in terms of classical electro-magnetic waves. 
\vspace{1mm}
\end{abstract} 
%
%
%
%
%
%
\vspace{0.5cm}
\newpage
\setcounter{page}{1}
\seqnoll
\bc{
\section{\large Introduction}
\label{sec:introuction}
}\ec
%
%

\begin{figure}[htb]  
\vspace{-4cm}
\centerline{\includegraphics[width=18cm,angle=0]{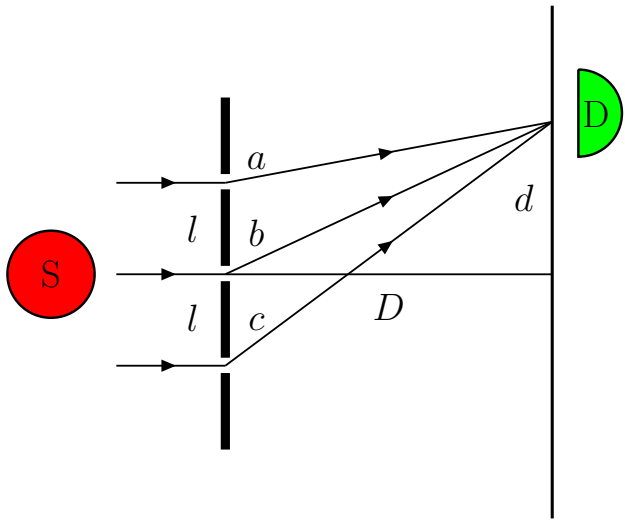}}
\vspace{-13cm}                 
\caption{
\label{fig:3slit} A three-slit Young interference setup. The normal mode annihilation operators $a$, $b$, and  $c$ also denote the various thin-slits. The source-mode annihilation operator is $s$.
The inter-slit distance is $l$ and $d$ denotes the position of the detectors. The distance $D$ between the Young interferometer and the detection plane is supposed to be large compared to any other length-scale.
}
\label{fig:2slit}
\end{figure}

Some time ago Sorkin \cite{Sorkin_1994} introduced a parameter $k_S$ defined for arbitrary complex numbers $\alpha$, $\beta$, and $\gamma$,
\begin{equation}
\label{eq:sorkin_1}
 \kappa_S \equiv \frac{1}{P} (P_{\alpha\beta\gamma} - P_{\alpha\beta} - P_{\alpha\gamma} - P_{\beta\gamma}+ P_{\alpha} + P_{\beta}+ P_{\gamma})  ~ ,
\end{equation} 
with $P_{\alpha\beta\gamma}=|\alpha+ \beta + \gamma|^2$, $P_{\alpha\beta}=|\alpha+ \beta|^2$, and $P_{\alpha} =|\alpha|^2$ and similarly for other combinations. $P$ is a suitably chosen normalization in order to factor out possible unimportant constants. Here we put $P=1$  but include it when convenient. Sorkin observed that $\kappa_s = 0$ as a {\sl mathematical identity} for the arbitrary complex numbers $\alpha$, $\beta$, and $\gamma$.
Since the complex numbers in the definition of $\kappa_S$ can be interpreted as quantum-mechanical probability amplitudes for physical events, it was, nevertheless, argued that a non-zero value of $\kappa_S$   could be used as test of some of the fundamental ingredients of quantum mechanics, namely the superposition principle and  Born's rule for obtaining measurable probabilities from quantum mechanical probability amplitudes. A physical realization corresponding to the symbol $P_{\alpha\beta\gamma}$  could, e.g.,  correspond to the detection probability in a three-slit Young interferometer as illustrated in Fig.\ref{fig:3slit}.  With one of the slits ($\gamma$) closed, $P_{\alpha\beta}$ should then be identified with the corresponding detection probability, and  $P_{\alpha}$ should correspond to two slits ($\beta$ and $\gamma$) closed and similarly for other combinations of the probability amplitudes $\alpha$, $\beta$, and $\gamma$. 
Various theoretical and experimental oriented considerations of these  ideas of Sorkin have recently been  under  investigation \cite{Sinha_2010,Park_2012, Raedt_2012, Sollner_2012, Gagnon_2014, Savant_2014, Sinha_and_Sinha_2015, Magana_2016, Kauten_2017,Cotter_2017,Rengaraj_2018}.

It is now of crucial importance to specify the identification above in a clear physical manner when making use of one and the same  experimental setup with  a given source and detector system. It is, e.g., then not obvious that  closing one slit in a three-slit Young interferometer is physically equivalent to a  two-slit situation to be used in the  experimental determination of $\kappa_S$.  By imposing proper boundary conditions for  the various Young interferometer configurations,  it has actually been argued  that a non-zero value of $\kappa_S$ quite naturally  emerges \cite{Raedt_2012, Savant_2014, Sinha_and_Sinha_2015, Rengaraj_2018}. In very elementary terms, and focusing on a purely quantum field theoretical  treatment, we will confirm that this is the case. 
Furthermore, one may raise  questions on the  quantum-mechanical nature of the prepared source state. We will verify, what has been known for a long period of time, that the interference pattern in all the cases   we consider does not depend on the quantum nature of state of the source, at least if we consider mono-chromatic sources and a single photon detector. Apart from an overall factor, the interference pattern will therefore be the same  for a source prepared in, e.g.,  a quantum-mechanical Fock-state of photons or for  a conventional coherent state. As is well-known, the use of  coherent states naturally leads to the interference of classical electro-magnetic fields (see, e.g., Ref.\cite{SER_QED} and references cited therein). The observation of a non-zero value of $\kappa_S$ is therefore not exclusively related to quantum-mechanical interference effects. 


%
\bc{
\section{\large The Photon Detector}
\label{sec:glauber}
}\ec
%
%
%

  We first recall a simple quantum field theoretical treatment  of quantum interference effects in a two-slit Young interferometer (see, e.g., Refs. \cite{Walls_1977,Mandel_99}), illustrated in Fig.\hspace{0.25mm}\ref{fig:3slit}, for a mono-chromatic source with wave-number $k$ and angular frequency $\omega=ck$. The Glauber theory of quantum coherence \cite{glauber63} is then  used in order to find the corresponding probability for single photon-detection. Below this analysis will be extended to a three-slit Young interferometer configuration. For a  prepared quantum  state $|\psi \rangle$ of the source $S$, and for a properly designed detector,  the detection probability  of one photon, with the port $c$ closed,  is related to the absorption of a photon in the detector described by the process
\begin{equation}
\label{eq:sorkin_2}
  |\psi \rangle \rightarrow E^{(+)}({\bf r},t) |\psi \rangle ~ .
\end{equation} 
Here
\begin{equation} 
\label{eq:sorkin_3}
   E^{(+)}({\bf r},t) = {\cal E} \bigg ( \, a \, \frac{e^{i\phi_a}}{r_a} + b \, \frac{e^{i \phi_a}}{r_b} \, \bigg ) ~ , 
\end{equation}
is the positive frequency part of one of the components of the second-quantized electric  field observable   $E^{(+)}({\bf r},t)$ at the position ${\bf r}$ and time $t$ at the detector far from the interferometer. The field $E^{(+)}({\bf r},t)$ is expressed in terms of outgoing  normal-mode annihilation operators $a$ and $b$. In a more rigorous setting  one should make use of appropriate Greens functions for system which, however, would make the points we are addressing less transparent. Furthermore,  $\phi_a = \omega (t - \tau_a)$ and $\phi_b = \omega (t - \tau_b)$ are suitable phases expressed in terms of time-delays $\tau_a$ and $\tau_b$. ${\cal E}$ is a common amplitude for the $a$ and $b$ modes  and $r_a$, $r_b$ are the in-plane distances from the various openings of the interferometer to  the detector $D$.
%
%
%
%
%
  According to the  fundamental Born's rule, the probability for single photon detection at the detector $D$ is then, apart from unimportant constants,   proportional  to $P_{ab}$ given by
\begin{equation}
 \label{eq:sorkin_4}
     P_{ab} = \sum_f | \langle f| E^{(+)}({\bf r},t) | \psi \rangle |^2  = \langle \psi | E^{(-)}({\bf r},t) E^{(+)}({\bf r},t) | \psi \rangle ~ , 
\end{equation}
 where we sum over all possible photon states $|f\rangle$. The probability $P_{ab}$ may therefore be written in a well-known general form using Eq.(\ref{eq:sorkin_3}), i.e., 
\begin{eqnarray} 
\label{eq:sorkin_5}
     P_{ab } = |{\cal E}|^2 \, \langle \psi | \, \bigg ( a^*a  + a^* b e^{-i(\phi_a - \phi_b)} ~~  
        + ~ ab^*  e^{i(\phi_a - \phi_b)}   + b^* b \, \bigg ) | \psi \rangle ~ .  
\end{eqnarray}

We now consider a prepared Fock state $|\psi\rangle$ for the source, i.e., 
\begin{equation}  \label{eq:sorkin_6}
    |\psi \rangle = \frac{s^{*n}}{\sqrt{n!}}|0 \rangle = \frac{(a^*+b^*)^{n}}{ \sqrt{2^nn!} }|0 \rangle = \frac{1}{\sqrt{2^n}}\sum_{k=0}^n  \, \bigg ( \frac{ n! }{ k! \,  (n-k)! } \bigg )^{1/2} \, |k \rangle_a \otimes |n-k \rangle_b  \, ~ ,
\end{equation}
where $|k \rangle_a$ and $|n-k \rangle_b$ represent the Fock states of photons emerging from the  slits $a$ and  $b$.
Here we  expressed the initial state $ |\psi \rangle$ in the  $a$ and $b$ modes using  a   boundary condition  at the two identical thin-slits, i.e., 
\begin{equation} 
\label{eq:sorkin_7}
  s = \frac{1}{\sqrt{2}} \, \big ( a + b  \big ) ~ . 
\end{equation}
This relation  does not represent the result of a unitary transformation. However, by including an additional local source, with a  mode operator $s_V=(a-b)/\sqrt{2}$, we have  a unitary $U(2)$ transformation connecting the pair of independent mode operators $(s,s_V)$ and the independent mode operators $(a,b)$ \cite{gerry_knight_2005}.  The number operator of photons will then be conserved. Expressed in a somewhat different manner, fundamental  commutation relations for mode operators applied to a completely symmetric Young interferometer naturally leads to the condition Eq.(\ref{eq:sorkin_7}). In the discussion below on the three-slit Young interferometer two  easily constructed local source operators have to be included in a similar manner and a corresponding unitary $U(3)$ transformation can easily be found preserving the number of photons. If  quantum states of such local modes are present with, e.g., random phases, the visibility of interference patterns will in general  be diminished. In all of the considerations below we, however, assume that  the quantum states of  such local modes are the vacuum state. We can therefore  suppress their presence in the considerations below.

The state vectors in Eq.(\ref{eq:sorkin_6}) describe the  superposition of all  possible combinations that can occur with appropriate weights,  for photons passing  through the various slits  at the same time. We stress again that we only need the asymptotic form of the field $E^{(+)}({\bf r},t)$ at the detector and a relation like Eq.(\ref{eq:sorkin_7}) for the mode operators in order to complete the analysis for all relevant detection probabilities.
It now follows that 
\begin{equation} 
\label{eq:sorkin_8}
     \langle \psi | a^*a | \psi \rangle =  \langle \psi | b^*b | \psi \rangle=  \langle \psi | b^*a | \psi \rangle=  \langle \psi | a^*b | \psi \rangle= \frac{n}{2} ~ ,
\end{equation}
and therefore 
%
%
\begin{gather}     \label{eq:sorkin_9}
     P_{ab} = |{\cal E}|^2 \, \frac{n}{2} \, \bigg ( \, \frac{1}{r_a^2} + \frac{1}{r_b^2} 
     + \, \frac{2\cos( \phi_a - \phi_b )}{r_a r_b}  \,  \bigg ) ~ . 
\end{gather}
Similar expressions can be   obtained for the probilities $ P_{ac}$ and $P_{bc}$. For the  one-slit case  the approximations used above lead to $P_a =|{\cal E}|^2 n/ r_a^2$ and similarly  for $P_b$ and $P_c$.
%
%
%
%

With all slits open in Fig.\ref{fig:3slit},  we extend the discussion above with the asymptotic field $ E^{(+)}({\bf r},t)$ in Eq.(\ref{eq:sorkin_3})  replaced by
\begin{equation}
\label{eq:sorkin_10}
   E^{(+)}({\bf r},t) = {\cal E} \bigg ( \, a \, \frac{e^{i\phi_a}}{r_a} + b \, \frac{e^{i \phi_a}}{r_b} \, c \, \frac{e^{i \phi_c}}{r_c} \,\bigg ) ~ . 
\end{equation}
Correspondingly, the initial state $|\psi\rangle$ is  expressed in terms of the $a$, $b$, and $c$ mode operators, i.e., 
\begin{equation} 
\label{eq:sorkin_11}
    |\psi \rangle = \frac{(a^*+b^*+c^*)^{n}}{ \sqrt{3^nn!} } \, |0 \rangle = \frac{1}{\sqrt{3^n}}\sum_{k=0}^n \sum_{l=0}^n \, \bigg ( \frac{ n! }{ k! \, l! \, (n-k-l)! } \bigg )^{1/2} \, |k \rangle_a  \otimes|l \rangle_b \otimes|n-k-l \rangle_c ~ ,
\end{equation}
where we have made use of the  multi-nomial theorem. As in the two-slit case, Eq.(\ref{eq:sorkin_11}) describes the  superposition of the possible combinations that can occur with appropriate weights, for photons passing through different slits  at the same time.  It is straightforward to verify  that the extension of Eq.(\ref{eq:sorkin_8}) is given by 
\begin{gather}  \nonumber
     \langle \psi | a^*a | \psi \rangle =  \langle \psi | b^*b | \psi \rangle= \langle \psi | c^*c | \psi \rangle =  \langle \psi | a^*b | \psi \rangle=\langle \psi | b^*a | \psi \rangle  \\
\label{eq:sorkin_12}
 = \langle \psi | a^*c | \psi \rangle = \langle \psi | c^*a | \psi \rangle= \langle \psi | b^*c | \psi \rangle =
\langle \psi | c^*b | \psi \rangle =\frac{n}{3} ~ .
\end{gather}
%
%
%
%
%
 The  three-slit probability $P_{abc}$   is therefore given by 
\begin{gather}
\label{eq:sorkin_13}
     P_{abc} = |{\cal E}|^2 \, \frac{n}{3} \, \bigg ( \frac{1}{r_a^2} + \frac{1}{r_b^2} + \frac{1}{r_c^2} 
     +  \frac{2\cos( \phi_a - \phi_b )}{r_a r_b}  \ +  \frac{2\cos( \phi_b - \phi_c )}{r_b r_c}   +  \frac{2\cos( \phi_a - \phi_c )}{r_a r_c}  \bigg ) ~ . 
\end{gather}
%
%

\begin{figure}[htb]  
\vspace{-1.2cm}
\centerline{\includegraphics[width=18cm,angle=0]{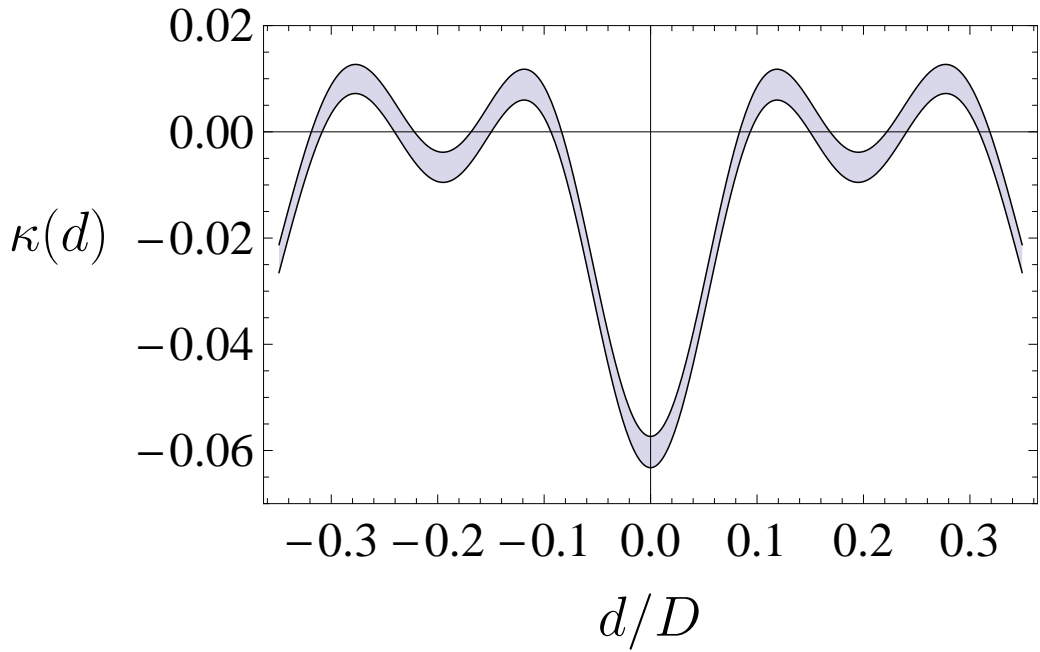}}
\vspace{-14.5cm}                 
\caption{The Sorkin parameter $\kappa_s \rightarrow \kappa(d)$, as defined in Eq.(\ref{eq:sorkin_18}) normalized by $P=P_{abc}(d=0)$, as a function of $d/D$. The parameter used are as in Ref.\cite{Rengaraj_2018} with $\lambda = 0.05$ m, $l=0.13$ m and $D=1.25$ m. For reasons of simplicity we neglect the width of the slots. The figures illustrates the sensitivity in reaching the  degeneracy value $\kappa(d)=0$ for a small range of the normalization constants $n_1$ and $n_2$ in  Eq.(\ref{eq:sorkin_18}). Here we make use of $n_2= 2/3 +1.3\%$. The upper curve corresponds to $n_1=1/3+ 1.3\%$ and the lower curve to  $n_1=1/3+ 1.2\%$.}
\label{fig:3slit_2}
\end{figure}

The various detection probabilities discussed above are all proportional to the number of photons $n$ of the initial state $|\psi \rangle$. In general all the results above will actually remain the same for any mono-chromatic initial quantum state, pure or mixed,  by replacing $n$ with the corresponding mean value $\langle n \rangle$. In order to verify this  fact we make use of Glaubers representation \cite{glauber63,sudarshan_etc_1963} for a general single-mode quantum state in terms of conventional coherent states $|\alpha \rangle$ \cite{skagerstam_klauder} namely
\begin{equation} \label{eq:sorkin_14}
     \rho = \int d^2\alpha \, {\cal P}(\alpha,\alpha^*) \, |\alpha \rangle \langle \alpha | ~ .
\end{equation}
A detection probability $P_D$ is then evaluated according to 
\begin{equation} \label{eq:sorkin_15}
     P_D \equiv \textrm{Tr} \big [ \, \rho \, E^{(-)}({\bf r},t) E^{(+)}({\bf r},t) \, \big ] =\int d^2\alpha \, {\cal P}(\alpha,\alpha^*) \, \langle \alpha |  E^{(-)}({\bf r},t)   E^{(+)}({\bf r},t) | \alpha \rangle ~ . ~ 
\end{equation}
We illustrate the procedure in terms of the two-slit slit configuration with $P_D=P_{ab}$. Expressing  coherent state $|\alpha \rangle =D(\alpha)|0\rangle$ in terms of the displacement operator $D(\alpha) = \exp( \alpha s^* - \alpha^* s )$,  and  by making use of the mode operator relationship  Eq.(\ref{eq:sorkin_7}), it is clear that
\begin{equation}
\label{eq:sorkin_16}
    |\alpha \rangle = \big | \frac{\alpha}{\sqrt{2}} \rangle_a \otimes \big | \frac{\alpha}{\sqrt{2}} \big \rangle_b   ~ .
\end{equation}
Since $|\alpha \rangle$ is an eigenstate of the observable $E^{(+)}({\bf r},t)$,  one easily finds the same expression for $P_{ab}$ as in Eq.(\ref{eq:sorkin_9}) with $n$ replaced by $\langle n \rangle \equiv \langle s^*s\rangle $  using
\begin{equation}
\label{eq:sorkin_17}
\frac{1}{2}\langle n\rangle \equiv \frac{1}{2}\textrm{Tr} \big [ s^*s\big] =\frac{1}{2}\int d^2\alpha \, {\cal P}(\alpha,\alpha^*) |\alpha|^2 = \langle a^*a\rangle =  \langle b^*b\rangle = \langle a^*b\rangle = \langle ab^*\rangle\, .
\end{equation}
Apart from the replacement of  $n$ with  $\langle n \rangle$,  the interference pattern exhibited by $P_{ab}$ then stays the same and does not depend on the details of the initially prepared quantum states of the source. The same reasoning applies to all detection probabilities considered above.

We now introduce the combination $\kappa (d)$,  motivated by Eq.(\ref{eq:sorkin_1}), as defined by
\begin{gather}
\kappa (d) \equiv  P_{abc} - n_2( P_{ab} + P_{ac} + P_{bc}) + n_1(P_{a} + P_{b}+ P_{c}) \,\, ,
\label{eq:sorkin_18}
\end{gather}
where  we have introduced two normalization parameters $n_1$ and $n_2$. With $n_1=n_2 =1$, the Sorkin parameter $\kappa_S$ and $\kappa(d)$ are, at least symbolically,   identical.  But then
\begin{gather}
 \kappa(d) = |{\cal E}|^2 \, \frac{\langle n \rangle}{3} \, \bigg ( \frac{1}{r_a^2} + \frac{1}{r_b^2} + \frac{1}{r_c^2} 
     -  \frac{\cos( \phi_a - \phi_b )}{r_a r_b}  \ -  \frac{\cos( \phi_b - \phi_c )}{r_b r_c}   -  \frac{\cos( \phi_a - \phi_c )}{r_a r_c}  \bigg ) \, ,
\end{gather}
is in general non-zero as a function of the position $d$ of the photon detector. Due to the two-slit conditions Eq.(\ref{eq:sorkin_8}) and three-slit conditions Eq.(\ref{eq:sorkin_12}) it is, however, clear that the parameters $n_1$ and $n_2$ has to be adjusted in order to have the same average number of photons passing through the various slit combinations under consideration. Without loss of generality, we should therefore use $n_1=1/3$ and $n_2=2/3$. The physical conditions are then the same for  the various slit configurations and one then finds that $\kappa (d) =0$. As an example, we illustrate  in Fig.\ref{fig:3slit_2} the sensitivity in the approach to the degeneracy point, defined by $\kappa (d)= 0$, for various choices of the normalization parameters $n_1$ and $n_2$. Other parameters used correspond to a recent experiment by Rengaraj et al. \cite{Rengaraj_2018}. We find it remarkable that we can reproduce some features  of Ref.\cite{Rengaraj_2018}  in view of the simplicity of the arguments put forward in these comments.

\bc{
\section{\large Remarks}
\label{sec:comments}
}\ec
%
%

In accordance with other considerations, as in Refs. \cite{Raedt_2012, Savant_2014, Sinha_and_Sinha_2015, Rengaraj_2018}, we have argued that one cannot make a straightforward physical identification of the quantum-mechanical probability amplitudes to be used in the definition of the Sorkin parameter $k_S$. As we have seen in the case of the one-, two-, and three-slit Young interferometer configurations, the intensity of the source considered has to be adjusted in order to describe the same physical conditions. It then follows that the corresponding identification of the Sorkin parameter $\kappa_S$ is naturally zero. If not properly adjusted  a non-zero value emerges without any contradiction with the basic rules of quantum mechanics. 

For a mono-chromatic source, the interference patterns discussed in these comments do not depend on the nature of the quantum state of the source and only the mean number of photons is of importance. This  has the consequence that there is no fundamental difference between classical and quantum-mechanical interference when making use of a single photon detector, a fact that is well established \cite{Grangier_1986}.
Even though the concept of a photon has been disputed \cite{Lamb_95},  the interference pattern as builded up by single-photon events can, with current technology, rather easily be demonstrated (see, e.g., Ref.\cite{Galvez_2014}) and, of course, agrees with the interference pattern as obtained in terms of classical optics.

The Feynman-path integral approach to quantum interference and the notion of non-classical paths for photons appears to play an important role in  many of the current discussions on the Sorkin parameter $\kappa_S$. For a non-relativistic particle the notion of a, not necessarily classical, path expressed in terms of co-ordinates makes much sense even though this has to be used with care (see, e.g., Ref.\cite{klauder_et_al_1984}). For highly relativistic particles the notion of a co-ordinate needs clarification  since one can argue that the components of a position observable for a massless particle with non-zero helicity, like a photon, are, due to topological reasons,  non-commuting \cite{skagerstam_1994}. In the quantum field theoretical approach to interference phenomena as discussed in these comments, such considerations do, however,  not play any role.
%
%
%
%
%
%
%
%
%
%
%
%

%
%
%
%
%
%
%
%
%
%
%
%
%
%
\begin{center}
   {\bf \large REFERENCES}
\end{center}
\noindent For the convenience of the readers we have attached a capital ${\bf R}$ to those
references which are reprinted in Ref.\cite{skagerstam_klauder}.
%
 
%
%
%
%
%
%
%

\begin{thebibliography}{99}
%
%
\bibitem{Sorkin_1994}  R. D. Sorkin, `{\sl Quantum Mechanics as Quantum Measure Theory~}'',  Mod. Phys. Lett. {\bf A9}, 3119 (1994).
%
%
\bibitem{Sinha_2010} U. Sinha, C. Couteau, T. Jennewein, R. Laflamme,  and G. Weihs,  ``{\sl Ruling Out Multi-Order interference in Quantum Mechanics~}'', Science {\bf 329}, 418  (2010).
%
%
 \bibitem{Park_2012} D. Park, K. Moussa, and R. Laflamme  ``{\sl Three Path Interference Using Nuclear Magnetic Resonance: A Test of the Consistency of Born's Rule~}'', New J. Phys. {\bf 14}, 113025 (2012).
%
%
\bibitem{Raedt_2012} H. De Raedt, K. Michielsen, and K. Hess, ``{\sl Analysis of Multipath Interference in Three-Slit Experiments~}'',  Phys. Rev.  A {\bf 85},  012101 (2012).
%
%
\bibitem{Sollner_2012} I. S\"{o}llner  et al.,  ``{\sl Testing Borns Rule in Quantum Mechanics for Three Mutually Exclusive Events~}'',  Found Phys. {\bf 42}, 742 (2012).
%
\bibitem{Gagnon_2014} E. Gagnon, C. D. Brown, and A. L. Lytle, ``{\sl Effects of Detector Size and Position on a Test of Born's rule Using a Three-Slit Experiment~}'', Phys. Rev. {\bf A90}, 013832 (2014).
%
%
\bibitem{Savant_2014} R. Sawant,  J. Samuel, A. Sinha, S. Sinha, and U. Sinha, ``{\sl Nonclassical Paths in Quantum Interference Experiments~}'', Phys. Rev. Lett. {\bf 113},  120406 (2014).
%
%
\bibitem{Sinha_and_Sinha_2015}  A. Sinha, A. H. Vijay, and  U. Sinha, ``{\sl On the Superposition Principle in Interference Experiments~}'', Scientific Reports {\bf 5}, 10304 (2015).
%
%
\bibitem{Magana_2016} O. S.  Maga\~{n}a-Loaiza et al. , ``{\sl Exotic Looped Trajectories of Photons in Three-Slit Interference~}'' Nat. Commun. {\bf 7}, 13987 (2016).
%
%
\bibitem{Kauten_2017} T. Kauten, R. Keil, T. Kaufmann, B. Pressl, \v{C}. Brukner, and G. Weihs,  ``{\sl Obtaining Tight Bounds on Higher-Order Interferences with a 5-Path Interferometer~}'', New J. Phys. {\bf 19}, 033017 (2017).
%
%
\bibitem{Cotter_2017} J. P. Cotter, C. Brand, C. Knobloch, Y. Lilach, O. Cheshnovsky, and M. Arndt,  ``{\sl In Search of Multipath Interference Using Large Molecules~}'' Sci. Adv. {\bf 3}, 1602478 (2017).
%
%
\bibitem{Rengaraj_2018} G. Rengaraj, U. Prathwiraj, S. N. Sahoo, R. Somashekhar, and U. Sinha, ``{\sl Measuring the Deviation from the Superposition Principle in Interference Experiments~}'', New J. Phys. {\bf 20},  063049 (2018). 
%
%
%
%
\bibitem{SER_QED} B.-S. K. Skagerstam,  K. E. Eriksson, and P. K. Rekdal, ``{\sl Causality in Quantum Field Theory with Classical Sources - Quantum Electrodynamics~}'', ArXiv:1801.09947v1 [quant-ph] 30 Jan 2018.
%
%
%
%
\bibitem{Walls_1977}
D. F. Walls,  ''{\sl A Simple Field Theoretic Description of Photon Interference~}'', Am. J. Phys. {\bf 45},  952 (1977).
%
%
\bibitem{Mandel_99} L. Mandel, ``{\it Quantum Effects in One-Photon and Two-Photon Interference~}'', Rev. Mod. Phys.   {\bf 71},  S274 (1999).
%
%
 %
\bibitem{glauber63} R. J. Glauber, ``{\sl Photon Correlations~}'', Phys. Rev. Lett. {\bf 10}, 84 (1963) ({\bf R});  ``{\sl The
Quantum Theory of Optical Coherence~}'', Phys. Rev. {\bf 130}, 252 (1963) ({\bf R});
``{\sl Coherent and Incoherent States of the Radiation Field~}'',
ibid. {\bf 131}, 2766 (1963) ({\bf R}),  ''{\sl Optical Coherence and Photon Statistics~}''
in: C. DeWitt, A. Blandin, C. C.-T. (Ed.), ''{\sl Quantum Optics and
Electronics~}'', Les Houches, p. 621, Gordon and Breach (New York, 1965); ''{\sl Quantum Theory of Optical Coherence~}'',
Wiley-VCH (Weinheim, 2017).
%
%
%
%
%
\bibitem{gerry_knight_2005} See, e.g., Section 5.3 in C. C. Gerry and P. L. Knight, ``{\sl Introductory Quantum Optics~}'', Cambridge University Press   (Cambridge, 2005).
%
%
%
\bibitem{sudarshan_etc_1963} E. C. G. Sudarshan, ``{\sl
Equivalence of Semiclassical and Quantum Mechanical Descriptions of Statistical
Light~}'', Phys. Rev. Lett. {\bf 10}, 277 (1963) ({\bf R}); 
J. R. Klauder, J. McKenna, and D. G. Currie, ``{\sl On "Diagonal"
Coherent-State Representations for Quantum-Mechanical Density
Matrices~}'', J. Math. Phys. {\bf 6}, 734 (1965);
C. L. Metha and E. C. G. Sudarshan, ``{\sl Relation Between Quantum
and Semi-Classical Description of Optical Coherence~}'', Phys.
Rev. {\bf 138}, B274 (1965);
J. R. Klauder, ``{\sl Improved Version of Optical Equivalence Theorem~}'',
Phys. Rev. Lett. {\bf 16}, 534 (1966);
J. R. Klauder and B.-S. K. Skagerstam, ``{\sl Generalized Phasespace
Representation of Operators~}'', J. Phys. A {\bf 40}, 2093 (2007).
%
%
\bibitem{skagerstam_klauder}{ J. R. Klauder and B.-S. K. Skagerstam, ``{\it Coherent States - Applications in Physics and Mathematical Physics~}'', World Scientific  (Singapore, 1985 and Beijing 1988); B.-S. K. Skagerstam, ``{\sl Coherent States - Some Applications in Quantum Field Theory and Particle Physics~}'' in ``{\sl Coherent States: Past, Present, and the Future~}'', p. 469, Eds. D. H. Feng, J. R. Klauder, and M. R. Strayer, World Scientific   (Singapore, 1994); J. R.  Klauder, ''{\sl The Current State of Coherent States~}'',  contribution to the  7th ICSSUR Conference, Boston, 2001  (arXiv:quant-ph/0110108).}
%
%
%
%
%
%
\bibitem{Grangier_1986}
P.\ Grangier, G.\ Roger,  and A.\  Aspect ,  ''{\sl Experimental Evidence for a Photon Anticorrelation Effect on a Beam Splitter: A New Light on Single-Photon Interference~}'', Europhys. Lett.  {\bf 1},  173 (1986);
 A. Aspect and P. Grangier, ``{\sl Wave-Particle Duality for
Single Photons~}'', {\it Hyperfine Interactions~} {\bf 37},  3 (1987);  P. Grangier,
G. Roger and A. Aspect, ``{\sl Experimental Evidence for a Photon Anticorrelation
Effect on a Beam Splitter: A New Light on Single-Photon Interferences~}'', 
Europhys. Lett., {\bf 1}, 173 (1986).
 %
%
%
\bibitem{Lamb_95} W. E. Lamb, Jr., ``{\it Anti-Photon~}'', Appl. Phys. B {\bf 60}, 77 (1995).
%
%
\bibitem{Galvez_2014} E. J. Galvez, ``{\sl Resource letter SPE-1: Single-Photon Experiments in the Undergraduate Laboratory~}'',  Am. J. Phys. {\bf 82}, 1018 (2014); R. S. Aspden, M. J. Padgett, and G. C. Spalding, ``{\sl Video Recording True Single-Photon Double-Slit Interference~}'', Am. J. Phys. {\bf 84},  671 (2016).
%
%
%
\bibitem{klauder_et_al_1984} J. R. Klauder, C. B. Lang, P. Salomonson, and B.-S. K. Skagerstam, ``{\sl Universality of the Continuum Limit of Lattice Quantum Field Theories~}'',  Z. Phys. C {\bf 26}, 149 (1984); 
A. L. Grimsmo, J. R. Klauder, and B.-S. K. Skagerstam, ``{\sl Anomalous Paths in Quantum Mechanical Path-Integrals~}'',  Phys. Lett. B {\bf 727}, 330 (2013).
%
%
%
\bibitem{skagerstam_1994} A. P. Balachandran, G. Marmo, and A. Stern, and B.-S. K Skagerstam, ``{\it Gauge Symmetries and Fibre Bundles - Applications to Particle Dynamics~}'',  Lecture Notes in Physics {\bf 188}, Springer Verlag  (Berlin 1983); 
B.-S. Skagerstam, ``{\it Localization of Massless Spinning Particles and the Berry Phase~}'', invited paper  in ``{\it On Klauders Path: A Field Trip - Festschrift for John R. Klauder on Occasion of His 60th Birthday~}'', p. 209,  Eds. G. G. Emch, G. C. Hegerfeldt, and L. Streit (World Scientific, 1994).
%
\end{thebibliography}
\end{document}